\documentclass[aps,twocolumn,longbibliography]{revtex4-1}
\usepackage{amsmath, amssymb}
\usepackage{amsbsy, latexsym, amsfonts, amsthm}
\usepackage{mathtools}
\usepackage[colorlinks,citecolor=blue,linkcolor=blue,urlcolor=blue]{hyperref}

\begin{document}

\title{Considerations on Quantum Gravity Phenomenology}

\author{Carlo Rovelli}
\affiliation{Aix Marseille University, Universit\'e de Toulon, CNRS, CPT, 13288 Marseille, France,\\ 
Perimeter Institute, 31 Caroline Street North, Waterloo, Ontario, Canada, N2L 2Y5,\\ 
The Rotman Institute of Philosophy, 1151 Richmond St.~N London, Ontario, Canada, N6A 5B7.}

\begin{abstract}
\noindent 
I describe two phenomenological windows on quantum gravity that seem promising to me. I argue that we already have important empirical inputs that should orient research in quantum gravity.
\end{abstract}

\maketitle  

I do two things in this brief note.   First, I describe the two directions towards quantum gravity phenomenology that seem more promising to me.   Then, I list some considerable empirical information that we have obtained lately, which I think is relevant for understanding the quantum properties of gravity, and I believe are unwisely disregarded.  This is also the opportunity for some general considerations on the~topic. 

\section{Where Are We Going to See Quantum Gravity~Effects}

\subsection{Gravity-Induced~Entanglement}

It is possible to probe a plausible and genuine quantum gravity effect in the laboratory with~technology that is not far from the one available today.  Surprisingly, nobody had realized that this was  the case until~a few years ago.  The~trick that makes this possible is that this is a (genuine, but) non-relativistic quantum gravitational~effect.

Here is the main idea~\cite{Bose2017a,Marletto2017a} (for related ideas, see~\cite{Howl2020}).  Two systems, $A$ and $B$, each with mass $m$, are each put into the quantum superpositions of two different positions, say, $L$ and $R$.    This generates a state formed by four branches:
\begin{eqnarray}
&&(|R\rangle_A+|L\rangle_A)\otimes (|R\rangle_B+|L\rangle_B)=\nonumber\\
&& \hspace{1cm}
|R,R\rangle+|R,L\rangle+|L,R\rangle+|L,L\rangle.
\label{1}
\end{eqnarray}

The systems are arranged in such a way that in one of these four branches, the two masses are at a small distance $d$ from each other, and~they are kept so for a time $t$. Then, the components of each of the two systems are recombined. 

The vicinity of the masses in one of the branches generates a gravitational interaction. This has the effect of altering the evolution of the phase of the branch.   In~a relativistic picture, this is because the gravitational field is different in each of the branches: the gravitational field is in a superposition of classical configurations; in the branch where the particles are close, each particle feels  the time dilatation due to the vicinity of the other mass~\cite{Christodoulou2018c}.  In~the non-relativistic picture, the~same effect is interpreted as due to gravitational potential energy $V=-Gm^2/d$.  Since the phase evolves with the energy $H$ as in $exp\{-iHt/\hbar\}$, the~total change in the phase of that branch with respect to the others is then clearly
\begin{equation}
\delta \phi = \frac{G m^2 t}{\hbar d}. 
\end{equation}

The change of phase in one branch has the effect of entangling the two systems, which, as~\eqref{1} shows, were not entangled to start with. The~fact that they are entangled can then be tested in the~lab. 

The crucial observation is that today's technology is not far from the possibility of keeping nano-particles in a superposition and at a distance $d$ from each other for a time $t$, such that $\delta\phi\sim\pi$ \cite{Bose2017a}. Hence, if~the gravitational field can be in a superposition, the~effect follows. Since we know from general relativity that the gravitational field is the same entity as the geometry of spacetime, the~measurement of this effect amounts to detecting an effect that follows from the superposition of spacetime~geometries. 

The power of this setup is in fact even stronger. The~reason is a well-known fact in quantum information: it is not possible to entangle two quantum systems by having them both interact with a third classical system.  In~this setup, the~two systems are $A$ and $B$, and~the third system is the gravitational field. If~we find $A$ and $B$ entangled by the gravitational interaction, then the gravitational field cannot be classical~\cite{Marletto2017a}. 

To be sure, the~knowledge that gravity is mediated by a field (in fact, a relativistic field) is needed for the interpretation of the experiment.  If~gravity was an instantaneous action at a distance and not mediated by a field, then we could not conclude anything from the experiment itself.   Hence, the~subtlety at the basis of this experiment is that it can be performed in a non-relativistic regime, but~its full implication requires the knowledge (that we have) that gravity is mediated by a relativistic field. In~other words, a~positive outcome of the experiment is not compatible with a description of gravity as the result of a classical~field. 

When successfully performed, the~importance of this experiment will be major. It could well be the first clear manifestation of the fact that spacetime geometry is not~classical. 

Since it is a non-relativistic regime, the experiment would not differentiate current tentative theories of gravity (such as loop quantum gravity, string theory, asymptotic safety, or~others). All current tentative theories predict it.  It would rule out, on the other hand,  speculations such as those exploring the (very unlikely) possibility that gravity is not quantized, or that there is a gravitationally induced physical collapse of the wave function. Variants of the experiment have been proposed that might actually access the relativistic regime and test the discreteness of proper time~\cite{Christodoulou2018b,Christodoulou2020}, but these require a much higher experimental sensitivity. 

In the past, there have been numerous other ideas on testing hypothetical quantum gravity effects, but---as far as I could understand---none considered a plausible effect; namely, an effect actually predicted by the current credible quantum gravity theories.  The~gravity-induced entanglement experiment does~so. 

This, I believe, is a general point. I find that there is a common false impression that since quantum gravity is an open problem, then ``everything is possible" and any wide speculation can be counted as a ``possible'' quantum gravity phenomenon.  This is not good science, in my opinion.  Quantum gravity is an open problem because no quantum gravity effect has been measured yet, because~there are a few competing theories about what exactly happens at the Planck scale, and~because we do not have a way of empirically probing them. But~all these theories are expected to give the same indications about what does or does not happen at lower scales.  

As~always in science, a~priori everything is possible, but~there is a profound difference between an implausible wild speculation and  the predictions of a plausible, coherent~framework. This is a distinction a bit too much disregarded in today's fundamental physics, in my opinion.

\subsection{Dark Matter as Quantum Gravity Stabilized White~Holes}

The first black hole signal was detected long before any black hole signal was recognized as such. In~fact, a~strong radio signal from Sagittarius A*, the~gigantic black hole at the center of our galaxy, has been detected by radio antennas since the dawn of radio astronomy, without~people suspecting it could be due to a black~hole. 

It might be the same with quantum gravity.  Dark matter is a major unclear phenomenon~\cite{Adam2016}.  There are many candidate theories for explaining dark matter, virtually all of which require the hypothesis of new physics.    But~there is also a possibility that dark matter could be explained without any recourse to new physics (which makes this hypothesis more, not less, interesting). The~possibility is that dark matter might be formed by long-living Planck-size remnants of evaporated black holes. The~black holes could have been formed in the early universe, or~alternatively, if~the Big Bang was a Big Bounce, they might have crossed the~bounce. 

The idea of black hole remnants is an old one, recently revived by quantum gravity calculations that provide them with a realistic model: white holes with a large interior and a small horizon stabilized by quantum gravity~\cite{Bianchi2018e}.    A~large body of theoretical research converge today, indicating that spacetime can be continued past the central singularity of a black hole and into an anti-trapped region, namely, a white hole. The~singularity  itself is replaced by a quantum region where the Einstein equations are briefly~violated.  

 Macroscopic white holes are unstable because they can easily re-collapse into black holes, but~Planck-size ones are stabilized by quantum theory~\cite{rovelli2018small}.  White hole remnants need to be long-lived because the information they store needs a long time to exit, in~the form of low-frequency~radiation. 

This scenario is attractive, difficult to falsify, but~also hard to confirm. In this article, I do not cover the current work that explores its phenomenology~\cite{Vidotto2018a}. What  I intended to point out is that it might (well) be that we are already seeing a massive quantum gravity effect: dark~matter.

\section{What Do We Already Know about Quantum Gravity?}

I come to the second topic---results that are relevant to quantum gravity, which are already providing us with crucial~information. 

\subsection{Lorentz~Invariance}

The breaking of Lorentz invariance at the Planck scale may simplify the construction of a quantum theory of gravity~\cite{Horava29}.  This observation sparked a large theoretical enthusiasm for Lorentz-breaking theories some time ago, and~rightly so. But~that bubble of enthusiasm has been deflated by empirical observations.  A~large campaign of astrophysical observations has failed to reveal the Planck-scale breaking of the Lorentz invariance in~situations where it would have been expected if this track for understanding quantum gravity had been the good one~\cite{Liberati2013}. 

A methodological consideration is important at this point.  Popperian falsifiability is an important demarcation criterium for scientific theories (that is, if a theory is not falsifiable, we better not call it ``science''); however,~Popperian falsification  is rarely the way theories gain or lose credibility in~science.   

The way scientific theories gain or loose credibility in real science  is rather through a Bayesian gradual increase or decrease of the positive or negative confirmation from empirical data.  That is, when a theory predicts a novel phenomenon and we this to be right, our confidence in the theory grows; when it predicts a novel phenomenon and we do not find it, our confidence in the theory decreases.  Failed predictions rarely definitely kill a theory, because~theoreticians are very good at patching up and adjusting. But~failed predictions do make the success of a research program far less probable: we loose confidence in~it.   

Hence, this has been the effect of not finding  Lorentz violations in astrophysics: tentative quantum gravity theories that break Lorentz invariance might perhaps still be viable in principle, but~in practice, far fewer people bet on~them. 

\subsection{Supersymmetry}

What I wrote above is particularly relevant to the spectacular non-discovery of supersymmetry at the LHC~\cite{Canepa2019}.  While in the Popperian sense, the non-appearance of supersymmetric particles at the TeV scale does not rule out all the theories based on supersymmetry, including string theory, in~practice, the  strong disappointment of not finding what was expected counts heavily as a strong dis-confirmation, in~the Bayesian sense, of~all those~theories. 

People have written that the non-discovery of supersymmetry is a crisis for theoretical physics. This is nonsense, of course. It is only a crisis for those who bet on supersymmetry and string theory.   For~all the alternative theoretical quantum gravity programs that were never convinced by the arguments for low-energy supersymmetry, the~non-discovery of supersymmetry is not a crisis: it is a~victory. 

Precisely for the same reason that the discovery of supersymmetry would have been a confirmation of the ideas supporting the string supersymmetry research direction, the~non-discovery of supersymmetry at the LHC is a strong empirical indication against the search for quantum gravity in the direction of supersymmetric theories and~strings.  

Nature talks, and~we better~listen.

\subsection{Cosmological~Constant}

A case similar to the one above but even stronger concerns the sign of the cosmological constant.  The~cosmological constant is a fundamental constant of nature, part of the Einstein equations (since 1917), whose value had not been measured until recently.  An~entire research community has long worked, and is still working, under general hypotheses that lead to the expectation for the sign of the cosmological constant to be negative.  Even today, the~vast majority of the theoretical work in that community assume~it to be so.

Except that the sign of the cosmological constant is not negative.   It is positive, as~observation has convincingly shown~\cite{Adam2016}.

Once again, this counts as a strong dis-confirmation of the hypotheses on which a large community has worked in the past, and is still working on~today.

So far, we lack any direct evidence of a quantum gravitational phenomenon; however,~the non-detection of Lorentz violations around the Planck scale, the~non-discovery of super symmetric particles at the LHC, and the measurement of a positive cosmological constant are strong indications from Nature that disfavor the tentative quantum gravity theories that naturally imply these~phenomena.


\vspace{6pt}

\begin{thebibliography}{999}

\bibitem{Bose2017a}
Bose, S.;~Mazumdar, A.;~Morley, G.W.; Ulbricht, H.; Toro{\v{s}}, P.M.M.;  Geraci, A.A.;  Barker, P.F.;  Kim, M.S.; Milburn, G.
  {Spin Entanglement Witness for Quantum Gravity}. {\em Phy. Rev.
  Lett.} {\bf 2017}, \emph{119}, 240401,
 {{arXiv:1707.06050}}.

\bibitem{Marletto2017a}
Marletto, C.; Vedral, V. {Witness gravity's quantum side in the lab}.
{{\em Nature} {\bf 2017}, \emph{547}, 156--158}. http://dx.doi.org/10.1038/547156a.

\bibitem{Howl2020}
Howl, R.;~Vedral, V.;~Naik, D.;~Christodoulou, M.;~Rovelli, C.; Iyer, A.
  {Non-Gaussianity as a Signature of a Quantum Theory of Gravity}. {\em arXiv} {\bf 2021},
  {{ arXiv:2004.01189}}.

\bibitem{Christodoulou2018c}
Christodoulou, M.; Rovelli, C.~{On the possibility of laboratory evidence
  for quantum superposition of geometries}. {\em Phys. Lett. B}
  {\bf 2018}, \emph{792}, 64--68,
{{arXiv:1808.05842}}.

\bibitem{Christodoulou2018b}
Christodoulou, M.; Rovelli, C.~{On the possibility of experimental
  detection of the discreteness of time}. \emph{arXiv} \textbf{201}8,
{{ arXiv:1812.01542v1}}.

\bibitem{Christodoulou2020}
Christodoulou, M.;~Di Biagio, A:~Martin Dussaud, P.~{
An experiment to test the discreteness of time}
{{ arXiv:2007.08431}}. 

\bibitem{Adam2016}
Adam, R. {Planck 2015 results: I. Overview of products and scientific
  results}. {\em Astron. Astrophys.} {\bf 2016}, \emph{594}, 1--38.
{{arXiv:1502.01582}}.

\bibitem{Bianchi2018e}
Bianchi, E.;~Christodoulou, M.;~Ambrosio, F.D.;  Haggard, H.M.; Rovelli, C.
 {White holes as remnants: A surprising scenario for the end of a black
  hole}. {\em Class. Quantum Grav.} {\bf 2018}, \emph{35}, 225003.
  {{arXiv:1802.04264}}.

\bibitem{rovelli2018small}
Rovelli, C.; Vidotto, F. {Small black/white hole stability and dark
  matter}. {\em Universe} {\bf 2018}, \emph{4}, 127.
{{arXiv:1805.03872}}.

\bibitem{Vidotto2018a}
Vidotto, F. {Quantum insights on Primordial Black Holes as Dark Matter}. 
 {\em PoS} {\bf 2018}, {\em EDSU2018}, 046. {{arXiv:1811.08007}}.

\bibitem{Horava29}
Horava, P.~{Quantum gravity at a Lifshitz point}. {\em Phys. Rev. D  
  Part. Fields  Grav. Cosmol.} {\bf 2009}, \emph{79}, 084008. {{ arXiv:0901.3775}}.

\bibitem{Liberati2013}
Liberati, S. {Tests of Lorentz invariance: A 2013 update}. {\em Class.
 Quantum Grav.} {\bf 2013}, \emph{30}, 133001.
{{arXiv:1304.5795}}.

\bibitem{Canepa2019}
Canepa, A.~{Searches for supersymmetry at the Large Hadron Collider}.
{{\em Rev. Physics}
  {\bf 2019}, \emph{4}, 100033}. \linebreak http://dx.doi.org/10.1016/j.revip.2019.100033.


\end{thebibliography}
\end{document}